\documentclass[12pt]{article}
\usepackage{amsmath,amssymb,amsthm}
\usepackage[margin=1in]{geometry}
\usepackage[round]{natbib}
\usepackage{setspace}
\newtheorem{theorem}{Theorem}
\newtheorem{lemma}{Lemma}
\newtheorem{proposition}{Proposition}
\newtheorem{corollary}{Corollary}

\newcommand{\E}{\mathbb{E}}
\newcommand{\R}{\mathbb{R}}
\newcommand{\hrho}{\hat\rho}
\newcommand{\hpi}{\hat\pi}
\newcommand{\ha}{\hat a}
\newcommand{\Spp}{\Sigma_{\pi\pi}}

\usepackage[colorlinks=true,linkcolor=blue,citecolor=blue,urlcolor=blue]{hyperref}

\makeatletter
\let\SavedMaketitle\maketitle
\let\Saved@maketitle\@maketitle
\let\SavedTitle\title
\let\SavedAuthor\author
\let\SavedDate\date
\let\SavedAnd\and
\makeatother

\title{AI in Econometric Theory: An Example}
\author{Isaiah Andrews \and Anna Mikusheva\thanks{We thank Jiafeng Chen, Paul Goldsmith-Pinkham, Ashesh Rambachan, Jonathan Roth, and Jesse Shapiro for helpful comments.}}

\date{\today}

\begin{document}

\pagenumbering{roman}
\maketitle
%
%
%

\begin{abstract}
This note contains a story and a paper.  The paper is about the admissibility of two-stage least squares with weak instruments, a topic we have worked on together since 2019.  The story is about how we (or, rather, our LLMs) wrote the paper, and some thoughts on what this may mean for our field.
\end{abstract}

Since 2019 we have worked on and off on the joint paper ``GMM is Inadmissible Under Bounded Identification Strength'' \citep{AndrewsMikusheva2024GMM}.  This paper started from the observation that common GMM estimators are discontinuous at data realizations where the sample objective has a non-unique minimizer, and the intuition that in settings with weak identification one should thus be able to do better with alternative estimators.
Motivated by this intuition, we attempted to prove that GMM estimators are dominated in the sense that there exist alternative estimators that are weakly better under all data-generating processes and strictly better under some.  
We were able to establish some results in this spirit, but subject to important caveats, namely that (a) the quantity of interest and the strength of identification were bounded in appropriate senses and (b) we could not analytically characterize a dominating estimator.  We submitted the paper to journals for potential publication, but it was rejected multiple times due in part to these limitations.\footnote{We received a revise and resubmit in 2024, though we have been negligent in turning it around.}  Both before and after these rejections, we tried to establish stronger results for the special case of two-stage least squares under squared error loss, but repeatedly failed to either relax our conditions or analytically characterize a dominating estimator.

Motivated by the progress of AI in math \citep[e.g.,][]{Alpoge2026,RamosHulakQueiroz2026,OpenAI2026NavierStokes,OpenAI2026SharingMath}, we recently experimented with using a commercially available frontier AI model (specifically, GPT-6 Astra) to try to solve these questions for two-stage least squares with squared error loss.  We found that the model could (a) prove that two-stage least squares is dominated without the boundedness conditions our proofs imposed and moreover (b) provide a variety of closed-form expressions for dominating estimators.  Indeed, the model could go further, proving dominance in a more general setting that nests models with heterogeneous treatment effects, where the target parameter is defined precisely as the two-stage least squares estimand, a more demanding setting for dominance results.\footnote{The current generation of models was not able to solve some follow-up questions we posed.  In particular, the models did not succeed in analytically characterizing an estimator which dominates two-stage least squares and is itself undominated by anything else.  We plan to return to that question after future model releases, and are particularly interested in whether such an alternative estimator might produce larger performance gains in empirically relevant settings than those found below.} We then used a combination of GPT-6 Astra and Claude Opus 5.5 to expand the results into the note below, which is entirely AI-written, though we asked for revisions in some places and take full responsibility for any errors.\footnote{We have both carefully reviewed the note and proofs, in addition to using AI for verification. A Lean formalization is available \href{https://economics.mit.edu/sites/default/files/2026-10/TSLS\%20Lean\%20Formalization.zip}{here}.}  Were we to write the note ourselves, we would make a number of further edits for clarity, emphasis, and style, but we provide it as-is in the spirit of illustrating current capabilities.

Given recent AI-driven results in math, that AI is now capable of proving what was, to us, a difficult result is entirely unsurprising.  Nevertheless, it is a striking change from our own experience with AI models in the relatively recent past: for instance, despite sustained experimentation, neither of us found them particularly helpful for our own proof writing prior to early 2026 (though we found them helpful for proof checking starting in 2025).  We think recent advances in capabilities, and the further advances which appear likely in the near future, call for a serious and ongoing re-evaluation of how our field assesses research outputs and allocates resources going forward.  We hope that, by talking about our own experiences as established researchers in the field, we might both encourage others to do the same and help contribute to what will no doubt be a much wider discussion on these and related questions across fields.

Before moving to the AI-written note, we want to be clear that we do not view this advance in capabilities as a tragedy.  A big part of why we each chose to work in econometrics is that we think the statistical methods used in economics research matter for the quality of the answers obtained, which are in turn consequential for a variety of decisions by both private and public actors.  Given that belief, we also think that better tools for producing knowledge in econometrics (and in a wide variety of other domains) are a very good thing, albeit one which demands appropriate institutional and policy responses.  Acknowledging the changes which have already taken place seems a useful early step towards formulating such responses.

\clearpage
\pagenumbering{arabic}
\setcounter{section}{0}
\setcounter{subsection}{0}
\setcounter{equation}{0}
\setcounter{footnote}{0}
\setcounter{table}{0}
\setcounter{figure}{0}
\setcounter{theorem}{0}
\setcounter{lemma}{0}
\setcounter{proposition}{0}
\setcounter{corollary}{0}

\makeatletter
\global\let\maketitle\SavedMaketitle
\global\let\@maketitle\Saved@maketitle
\global\let\title\SavedTitle
\global\let\author\SavedAuthor
\global\let\date\SavedDate
\global\let\and\SavedAnd
\makeatother

\title{Two-Stage Least Squares is Inadmissible with Weak Instruments}
\author{}
\date{}
\maketitle


\begin{abstract}
\noindent Two-stage least squares (TSLS) is the standard estimator in linear instrumental variables regression, and with weak instruments its behavior is poorly described by conventional approximations. We study the Gaussian limit experiment with known covariance and unrestricted reduced-form and first-stage coefficients, which accommodates heterogeneous treatment effects, and take the TSLS estimand as the target. TSLS is known to have infinite mean squared error when there are fewer than three instruments, and we show that it is inadmissible under squared error loss whenever there are at least three, with no restriction on identification strength. The argument is constructive: moving the TSLS estimate toward a bounded statistic when the observed first stage is close to zero lowers risk at every parameter value, and we give two simple dominating estimators with explicit tuning. In simulations calibrated to published specifications, the variance reductions when the reduced form and first stage are zero are large with three instruments but negligible with 21 or more, and the improvements in mean squared error at the estimated coefficients are small and concentrated in the specifications with the smallest first-stage Wald statistics.
\end{abstract}

\section{Introduction}\label{sec: intro}

Two-stage least squares (TSLS) is the standard estimator in linear instrumental variables (IV) regression. When instruments are weak, in the sense that the first stage is small relative to sampling uncertainty, conventional approximations to the behavior of TSLS are unreliable \citep{StaigerStock1997,StockWright2000}; see \citet{AndrewsStockSun2019} for a review. With normal errors, TSLS has finite moments only up to the degree of overidentification \citep{Kinal1980}, so its mean squared error is finite only when there are at least three instruments. With fewer instruments, any estimator with finite second moments dominates TSLS under squared error loss.\footnote{The requirement of at least three instruments recalls the inadmissibility of the sample mean of a normal vector in three or more dimensions \citep{Stein1956,JamesStein1961}. Here $k\ge3$ is the condition for TSLS to have finite risk.}

A decision-theoretic literature studies estimation with weak instruments. \citet{Chamberlain2007} uses invariance and minimax arguments to derive estimators and confidence sets in a linear IV model with a single structural coefficient. \citet{HiranoPorter2015} show that no estimator can be mean-unbiased, quantile-unbiased, or translation equivariant (in their distributional sense) when the first-stage coefficients are unrestricted near zero, while \citet{AndrewsArmstrong2017} construct an unbiased estimator when the sign of the first stage is known. Closer to our approach, other work improves on TSLS by shrinkage, under different criteria and asymptotic frameworks. \citet{Hansen2017} shows that a weighted average of TSLS and ordinary least squares, with weight determined by a Hausman statistic, has lower asymptotic risk than TSLS, under a particular weighted quadratic loss and local exogeneity, when there are more than two endogenous regressors. \citet{Spiess2017} shows that combining James--Stein shrinkage of the first stage with a control-function second stage yields an estimator that dominates TSLS in bias, with one endogenous regressor and at least four instruments. \citet{AndrewsMikusheva2022} derive the limit experiment for weakly identified generalized method of moments (GMM) models and study optimal decision rules there. Most directly related, \citet{AndrewsMikusheva2024GMM} show that GMM estimators, including TSLS, are inadmissible in this limit experiment under any bounded, uniformly convex loss when identification strength is bounded. Their argument uses a complete class theorem \citep{Brown1986} to show that admissible estimators must be uniformly continuous in the sample moments, which GMM estimators are not.

While this establishes that TSLS can be improved, it leaves open how. As \citet{AndrewsMikusheva2024GMM} write, ``Our proof for inadmissibility is non-constructive, in the sense that it does not deliver a dominating estimator.'' Their result also requires a bounded loss and a bound on identification strength. It therefore does not cover squared error loss on an unrestricted parameter space, under which TSLS has finite risk whenever there are at least three instruments.

A second gap concerns heterogeneous treatment effects. With multiple instruments and heterogeneous effects, TSLS estimates a weighted average of local average treatment effects under the assumptions of \citet{ImbensAngrist1994} \citep{ImbensAngrist1994,AngristImbens1995}. The constant-effect restriction that the reduced-form coefficients are proportional to the first-stage coefficients then generally fails. In the linear IV model, the decision-theoretic results discussed above impose this restriction (more generally, they assume correctly specified moment conditions). The weak-instrument-robust tests of \citet{AndersonRubin1949}, \citet{Kleibergen2002}, and \citet{Moreira2003} are likewise derived for the constant-effect model. When heterogeneity makes the reduced-form coefficients nonproportional to the first-stage coefficients, the null hypotheses these tests are built around fail for every value of the coefficient, so the tests do not target the TSLS estimand. \citet{AndrewsStockSun2019} note more broadly that existing weak-instrument-robust procedures assume constant effects. Recent work has begun to address this gap. \citet{Lee2018} gives variance estimators for TSLS when instruments identify different local average treatment effects, under strong instruments, and \citet{EvdokimovKolesar2019} study inference with many instruments and heterogeneous effects. \citet{Yap2025} and \citet{BootNibbering2024} develop inference that is robust to many weak instruments under heterogeneity, and \citet{Vestre2026} develops weak-instrument-robust tests for the TSLS estimand with a fixed number of instruments. \citet{KleibergenZhan2025} develop tests robust to both weak identification and misspecification for the pseudo-true value that minimizes the population continuously updated GMM objective, a target that with heterogeneous effects generally differs from the TSLS estimand. Our results concern estimation rather than inference, and apply to the TSLS estimand under heterogeneous effects with a fixed number of possibly weak instruments.

This note studies estimation of the TSLS estimand in the Gaussian limit experiment for linear IV. The covariance of the reduced-form and first-stage estimates is known, and we impose no restriction linking the reduced-form and first-stage coefficients, so the analysis allows for heterogeneous effects. We ask whether TSLS can be improved under squared error loss, at every parameter value, without any bound on identification strength. We show that it can, for every number of instruments $k\ge3$, every covariance matrix, and every positive definite weight matrix. Since the constant-effect model is a special case of our parameter space, the same conclusion holds there, for the structural coefficient.

Our first result is a local stabilization argument. When the first-stage Wald statistic is small enough, replacing the TSLS estimate by a bounded statistic lowers conditional risk at every parameter value. Consequently, any estimator that moves TSLS toward this statistic on that event, and only there, by a fraction that depends only on the Wald statistic, strictly dominates TSLS. The stabilization region is small, so these estimators coincide with TSLS except on an event of probability at most $\Pr(\chi^2_k\le1/2)$, about 0.08 when $k=3$. We further show that every estimator dominating TSLS must satisfy a calibration condition when the first stage is uninformative, and that some admissible estimator dominates TSLS, although our existence argument does not characterize it.

Our second result studies two simple dominating estimators, a truncated-linear and an exponential shrinkage of the TSLS estimate toward the same bounded statistic. Each has an explicit tuning condition, and we obtain sharper tunings by verifying finite sets of sufficient conditions in exact arithmetic. When $W$ is proportional to $\Spp^{-1}$, as for TSLS under homoskedasticity, we also bound the variance reduction at a zero reduced form and first stage available from any estimator of this form that dominates TSLS. Only the truncated-linear shape can attain this bound, which shrinks rapidly with the number of instruments. Neither estimator dominates the other in general, and we make no claim that either is admissible.

Finally, we evaluate these estimators in designs calibrated to the 70 specifications with at least three instruments in the sample of \citet{AndrewsStockSun2019}. When both the reduced form and the first stage are zero, the certified estimators reduce variance by about three quarters on average in the 30 specifications with three instruments, but negligibly in the 40 with 21 or more. At the estimated coefficients, by contrast, the improvements in mean squared error are below 0.1\% on average and at most 3.7\%, and they are concentrated in the two specifications with the smallest first-stage Wald statistics.

The main limitation of our analysis is that our dominance results are exact statements about the Gaussian limit experiment with known covariance. We do not establish risk dominance for feasible estimators that use estimated covariance and weight matrices, and since squared error loss is unbounded, such results would not follow from convergence to the limit experiment alone.

Section \ref{sec: setting} introduces the setting. Section \ref{sec: inadmissibility} shows that TSLS is inadmissible and characterizes dominating estimators. Section \ref{sec: dominators} develops the two simple dominators, and Section \ref{sec: simulations} reports simulations calibrated to published specifications.

\section{Setting}\label{sec: setting}

We study estimation in the Gaussian limit experiment for linear instrumental variables (IV) regression with weak instruments. Consider a scalar outcome, a scalar endogenous regressor, and $k$ instruments. Let $\hrho$ and $\hpi$ denote the coefficients from regressing the outcome and the endogenous regressor, respectively, on the instruments, after partialling out any exogenous controls. Under the weak-instrument asymptotics of \citet{StaigerStock1997}, the population coefficients shrink with the sample size at the root-$n$ rate, and suitably scaled estimates are asymptotically normal with a consistently estimable covariance matrix. Motivated by this approximation, we treat the scaled estimates as exactly normal with known covariance, and observe
\begin{equation}\label{eq: model}
\begin{pmatrix}\hrho\\ \hpi\end{pmatrix}\sim N\left(\begin{pmatrix}\rho\\ \pi\end{pmatrix},\Sigma\right),\qquad
\Sigma=\begin{pmatrix}\Sigma_{\rho\rho}&\Sigma_{\rho\pi}\\ \Sigma_{\pi\rho}&\Spp\end{pmatrix},
\end{equation}
where $\Sigma$ is positive definite. We place no restriction on the relationship between the reduced-form coefficients $\rho$ and the first-stage coefficients $\pi$, and take the parameter space to be $\rho\in\R^k$ and $\pi\in\R^k\setminus\{0\}$.

For a fixed positive definite weight matrix $W$, our target is
\begin{equation}\label{eq: estimand}
\theta(\rho,\pi)=\frac{\pi'W\rho}{\pi'W\pi},
\end{equation}
and the estimator we study is its plug-in analog, $\delta_0=\hpi'W\hrho/\hpi'W\hpi$. When $W$ is the second moment matrix of the instruments, $\delta_0$ is the two-stage least squares (TSLS) estimator and $\theta$ is the TSLS estimand. If treatment effects are constant and the linear IV model is correctly specified, then $\rho=\beta\pi$ and $\theta=\beta$. With heterogeneous effects and multiple instruments, by contrast, $\rho$ need not be proportional to $\pi$. Under the monotonicity assumption of \citet{ImbensAngrist1994}, TSLS then estimates a weighted average of local average treatment effects \citep{AngristImbens1995}, although with covariates this interpretation requires additional conditions on how they are specified \citep{BlandholBonneyMogstadTorgovitsky2026}. With multiple instruments this assumption is restrictive, and under the weaker partial monotonicity condition of \citet{MogstadTorgovitskyWalters2021} the weights can be negative. More generally, when the instrument-specific estimands are defined, $\theta$ is a linear combination of them with weights that sum to one \citep{Rotemberg1983,Andrews2019}, which \citet{GoldsmithPinkhamSorkinSwift2020} call Rotemberg weights. Leaving $\rho$ and $\pi$ unrestricted allows for all of these cases.

We evaluate an estimator $\delta=\delta(\hrho,\hpi)$ by its mean squared error, $R_{\rho,\pi}(\delta)=\E_{\rho,\pi}[(\delta-\theta(\rho,\pi))^2]$, which is defined for $\pi\neq0$. An estimator $\delta$ dominates $\delta'$ if $R_{\rho,\pi}(\delta)\le R_{\rho,\pi}(\delta')$ for all $(\rho,\pi)$, with strict inequality for some $(\rho,\pi)$, and strictly dominates $\delta'$ if the inequality is strict for all $(\rho,\pi)$. An estimator is admissible if no other estimator dominates it.\footnote{Under squared error loss, randomized estimators are weakly dominated by their conditional means given the data, so it is without loss to consider nonrandomized estimators.}

To understand the behavior of TSLS, it helps to purge the reduced-form estimate of its correlation with the first-stage estimate. Let $\Gamma=\Sigma_{\rho\pi}\Spp^{-1}$ and $\ha=\hrho-\Gamma\hpi$. Then $\ha\sim N(a,S)$ independently of $\hpi\sim N(\pi,\Spp)$, where $a=\rho-\Gamma\pi$ and $S=\Sigma_{\rho\rho}-\Sigma_{\rho\pi}\Spp^{-1}\Sigma_{\pi\rho}$. Since $(\ha,\hpi)$ is a one-to-one transformation of $(\hrho,\hpi)$, we can equivalently write estimators as functions of $(\ha,\hpi)$. Define
\[
b(v)=\frac{v'W\Gamma v}{v'Wv},\qquad \ell(v)=\frac{Wv}{v'Wv}.
\]
Substituting $\hrho=\ha+\Gamma\hpi$ and $\rho=a+\Gamma\pi$ shows that
\begin{equation}\label{eq: decomposition}
\delta_0=b(\hpi)+\ell(\hpi)'\ha,\qquad \theta(\rho,\pi)=b(\pi)+\ell(\pi)'a.
\end{equation}

Decomposition \eqref{eq: decomposition} splits TSLS into two parts. The first, $b(\hpi)$, is a ratio of quadratic forms in $\hpi$, and so is bounded, taking values in an interval $[b_-,b_+]$ determined by $W$ and $\Gamma$. When $\Sigma$ has the Kronecker form implied by homoskedasticity, $\Gamma$ is proportional to the identity and $b(\hpi)$ is constant, equal to the probability limit of ordinary least squares under weak-instrument asymptotics. The second part carries the information in the reduced form. Conditional on $\hpi$ it is normal with variance $\ell(\hpi)'S\ell(\hpi)$, which is of order $|\hpi|^{-2}$ and so explodes as the first-stage estimate approaches zero. Since $\E_{\rho,\pi}|\hpi|^{-2}$ is finite if and only if $k\ge3$, TSLS has finite risk at every parameter value if and only if $k\ge3$, consistent with the moment results of \citet{Kinal1980}. With $k\le2$ instruments any estimator with finite second moments dominates TSLS, so we assume $k\ge3$ throughout.

Our results depend on $\Sigma$ and $W$ through three constants. Let $K=\Spp^{1/2}W\Spp^{1/2}$, where $\Spp^{1/2}$ is the symmetric square root, and define
\[
\kappa=\frac{\lambda_{\max}(K)}{\lambda_{\min}(K)},\qquad
s_*=\lambda_{\min}\left(\Spp^{-1/2}S\Spp^{-1/2}\right),\qquad
H=\frac{b_+-b_-}{2}.
\]
The condition number $\kappa\ge1$ measures how far $W$ is from being proportional to $\Spp^{-1}$, with $\kappa=1$ for TSLS under homoskedasticity. The constant $s_*>0$ measures the reduced-form noise that remains after purging the first stage, relative to the first-stage noise. The half-width $H$ measures how much $b$ varies, and is zero under homoskedasticity.\footnote{All of our results remain valid if $\kappa$ and $H$ are replaced by upper bounds and $s_*$ by a positive lower bound. Such bounds can be certified exactly by rational matrix inequalities when $\Sigma$ and $W$ are rational.} Finally, let $Q=\hpi'\Spp^{-1}\hpi$ denote the Wald statistic for testing $\pi=0$, which equals $k$ times the first-stage $F$ statistic when the covariance is known. The distribution of $Q$ is noncentral $\chi^2_k$ with noncentrality $\pi'\Spp^{-1}\pi$.

\section{TSLS is Inadmissible}\label{sec: inadmissibility}

Decomposition \eqref{eq: decomposition} shows that TSLS adds to the bounded statistic $b(\hpi)$ a term whose conditional variance is of order $1/Q$. When the observed first stage is close to zero, this term contributes mostly noise. Our first result shows that, once $Q$ is small enough, replacing TSLS by $b(\hpi)$ lowers conditional risk at every parameter value.

\begin{lemma}\label{lem: replacement}
Let $\bar q=s_*/\{2(H+\sqrt{H^2+\kappa s_*})^2\}$. For all $\rho\in\R^k$, all $\pi\neq0$, and all $q\in(0,\bar q]$,\footnote{Here and below, conditional expectations given $Q=q$ are computed under the conditional distribution in which the direction of $\Spp^{-1/2}\hpi$ has density proportional to $\exp(\sqrt q\,\pi'\Spp^{-1/2}n)$ with respect to the uniform measure on the unit sphere. This is a version of the conditional distribution given $Q$ for every $q>0$.}
\[
\E_{\rho,\pi}\left[(b(\hpi)-\theta)^2-(\delta_0-\theta)^2\mid Q=q\right]\le-\frac{s_*}{2q}.
\]
\end{lemma}

Lemma \ref{lem: replacement} conditions on the observed Wald statistic, not on the parameters. The inequality holds for every $(\rho,\pi)$, including parameter values with a strong first stage and a large target, on the event that $Q$ falls below $\bar q$. Replacing TSLS by $b(\hpi)$ removes the term $\ell(\hpi)'\ha$, whose conditional variance is at least $s_*/Q$. It also changes the conditional bias. For targets $\theta$ near the range of $b$, the change in squared bias is bounded, and is outweighed by the variance reduction when $Q\le\bar q$. For targets far from that range, the squared bias of $b(\hpi)$ can be large. Averaging over the direction of $\hpi$, however, removing $\ell(\hpi)'\ha$ reduces the squared conditional bias as well once $Q$ is small. Appendix \ref{app: proofs} gives the proof.

Lemma \ref{lem: replacement} immediately yields a class of dominating estimators. For a measurable function $w:(0,\infty)\to[0,1]$, consider the estimator
\begin{equation}\label{eq: stabilized}
\delta_w=\delta_0-w(Q)\{\delta_0-b(\hpi)\}=b(\hpi)+\{1-w(Q)\}\ell(\hpi)'\ha,
\end{equation}
which moves TSLS toward $b(\hpi)$ by a fraction $w(Q)$ that depends on the observed first-stage strength.

\begin{theorem}\label{thm: stabilization}
Suppose $w(q)=0$ for $q>\bar q$, and $w$ is positive on a set of positive Lebesgue measure. Then for all $\rho\in\R^k$ and $\pi\neq0$,
\[
R_{\rho,\pi}(\delta_0)-R_{\rho,\pi}(\delta_w)\ge\frac{s_*}{2}\E_{\pi}\left[\frac{w(Q)}{Q}\right]>0.
\]
\end{theorem}

Since squared error is convex and $\delta_w$ is a convex combination of $\delta_0$ and $b(\hpi)$ with weights that depend only on $Q$, the conditional risk of $\delta_w$ given $Q$ is at most the corresponding combination of the conditional risks of $\delta_0$ and $b(\hpi)$. Theorem \ref{thm: stabilization} then follows from Lemma \ref{lem: replacement}. The estimator $\delta_w$ coincides with TSLS whenever $Q>\bar q$, so it modifies TSLS only when the observed first stage is very close to zero. Since $\bar q\le1/(2\kappa)\le1/2$ and the distribution of $Q$ is stochastically increasing in the noncentrality, $\delta_w$ differs from TSLS with probability at most $\Pr(\chi^2_k\le1/2)$, which is about 0.08 for $k=3$. The lower bound on the improvement does not depend on the reduced-form coefficients $\rho$, and depends on $\pi$ only through the noncentrality $\pi'\Spp^{-1}\pi$. It is positive at every parameter value but decays rapidly as the first stage strengthens, since the event $Q\le\bar q$ then becomes rare.

Theorem \ref{thm: stabilization} applies for any covariance matrix and weight matrix, and taking $w(q)=(1-q/\bar q)_+$ satisfies its conditions. It therefore implies the following.

\begin{corollary}\label{cor: inadmissible}
For every $k\ge3$, every positive definite $\Sigma$, and every positive definite $W$, TSLS is strictly dominated under squared error loss, and so is inadmissible. In particular, in the constant-effects model $\rho=\beta\pi$, with $\beta\in\R$ and $\pi\in\R^k\setminus\{0\}$, the estimator $\delta_w$ strictly dominates TSLS as an estimator of $\beta$.
\end{corollary}

The second part of Corollary \ref{cor: inadmissible} follows because the constant-effects model is a subset of our parameter space, on which $\theta=\beta$. \citet{AndrewsMikusheva2024GMM} show that GMM estimators, including TSLS, are inadmissible in this model under any bounded, uniformly convex loss when identification strength is bounded. Corollary \ref{cor: inadmissible} differs from their result for TSLS in three respects. First, it uses squared error loss, which is unbounded. Second, it places no bound on identification strength, or on $\beta$. Third, it is constructive, delivering an explicit dominating estimator. Their result, by contrast, covers general GMM estimators and a broad class of loss functions. The first part of Corollary \ref{cor: inadmissible} shows that TSLS remains inadmissible without constant effects, as an estimator of its own estimand.

Theorem \ref{thm: stabilization} exhibits one class of dominating estimators. Our next result shows that every estimator dominating TSLS shares a feature with it, which concerns its behavior when both the reduced form and the first stage are uninformative. Let $m=\E_{V\sim N(0,\Spp)}[b(V)]$ denote the mean of $b(\hpi)$ when $\pi=0$.

\begin{theorem}\label{thm: calibration}
Let $\delta$ be an estimator, written as a function of $(\ha,\hpi)$, with $R_{\rho,\pi}(\delta)\le R_{\rho,\pi}(\delta_0)$ for all $\rho\in\R^k$ and $\pi\neq0$. Then
\begin{enumerate}
\item $\E_{V\sim N(0,\Spp)}[\delta(u,V)]=m$ for Lebesgue-almost every $u\in\R^k$; and
\item $\E_{\rho,0}[\delta^2]\le\E_{\rho,0}[\delta_0^2]$ for every $\rho\in\R^k$.
\end{enumerate}
\end{theorem}

The first part of Theorem \ref{thm: calibration} requires that when the first stage is completely uninformative, a dominating estimator average to the same constant $m$ over the first-stage noise, whatever the value of the purged reduced form. TSLS satisfies this condition because $b$ is even and $\ell$ is odd, and so does any estimator of the form $b(\hpi)+f(\hpi)\ell(\hpi)'\ha$ with $f$ bounded and even, including $\delta_w$. The proof considers sequences of parameter values along which $\pi$ tends to zero while the target $\theta$ is held fixed at an arbitrary value. The risk comparison along these sequences restricts the mean of $\delta$ at $\pi=0$, and completeness of the normal location family then delivers the conclusion for almost every $u$. Although the target is undefined at $\pi=0$, the second part shows that a dominating estimator must also have second moments no larger than those of TSLS there.

Theorem \ref{thm: stabilization} shows that TSLS is inadmissible, but does not deliver an admissible estimator. Our final result in this section shows that an admissible estimator dominating TSLS exists. Let $\E_{0,0}$ denote expectations when $\rho=0$ and $\pi=0$.

\begin{theorem}\label{thm: admissible}
Let $\tilde\delta$ be an estimator that dominates $\delta_0$, and let $\mathcal{D}_{\tilde\delta}$ be the set of estimators $\delta$ with $R_{\rho,\pi}(\delta)\le R_{\rho,\pi}(\tilde\delta)$ for all $\rho\in\R^k$ and $\pi\neq0$. There exists an estimator $\delta^*\in\mathcal{D}_{\tilde\delta}$, unique up to almost-sure equality, that minimizes $\E_{0,0}[(\delta-m)^2]$ over $\mathcal{D}_{\tilde\delta}$, and $\delta^*$ is admissible. In particular, taking $\tilde\delta=\delta_w$ from Theorem \ref{thm: stabilization} yields an admissible estimator that strictly dominates TSLS at every parameter value.
\end{theorem}

The proof of Theorem \ref{thm: admissible} is a projection argument. By Theorem \ref{thm: calibration}, every element of $\mathcal{D}_{\tilde\delta}$ has a finite second moment when $\rho=0$ and $\pi=0$. Moreover, $\mathcal{D}_{\tilde\delta}$ is closed and convex in the corresponding $L^2$ space, so the minimizer exists and is unique. Any estimator that dominates $\delta^*$ also lies in $\mathcal{D}_{\tilde\delta}$. Taking limits along parameter sequences that approach $\rho=0$ and $\pi=0$ with the target held at $m$ shows that such an estimator has no larger value of the objective, so by uniqueness it coincides with $\delta^*$. It shows that admissible improvements on TSLS exist, but does not characterize them, and in particular does not imply that $\delta_w$, or the estimators in Section \ref{sec: dominators}, are themselves admissible.

\section{Two Simple Dominators}\label{sec: dominators}

Theorem \ref{thm: stabilization} shows that moving TSLS toward $b(\hpi)$ on the event $Q\le\bar q$, by a fraction that depends only on $Q$, improves on it. We next study two simple estimators of this type, each with a single tuning parameter. Both belong to the family
\begin{equation}\label{eq: taper}
\delta_f=b(\hpi)+f(Q)\ell(\hpi)'\ha,
\end{equation}
where $f:[0,\infty)\to[0,1]$. The estimator $\delta_f$ equals TSLS when $f\equiv1$ and $b(\hpi)$ when $f\equiv0$. Since $f(Q)$ is an even function of $\hpi$, every such estimator satisfies the necessary condition in Theorem \ref{thm: calibration}. Unlike TSLS, both estimators below have finite moments of all orders for every $k$, since $f(Q)\ell(\hpi)$ is bounded whenever $f(q)$ is bounded by a multiple of $q$ near zero.

The first estimator shrinks linearly toward $b(\hpi)$ when $Q$ falls below a cutoff $Q_*$:
\[
f_{c}(q)=\min\{1,q/Q_*\}.
\]
This is the estimator $\delta_w$ with $w(q)=(1-q/Q_*)_+$. By Theorem \ref{thm: stabilization}, it strictly dominates TSLS whenever $Q_*\le\bar q$, with $R_{\rho,\pi}(\delta_0)-R_{\rho,\pi}(\delta_{f_c})\ge(s_*/2)\E_\pi[(Q^{-1}-Q_*^{-1})_+]$. We refer to it as the truncated-linear estimator.

The second estimator shrinks toward $b(\hpi)$ by a factor that decays exponentially in $Q$:
\[
f_{e}(q)=1-\exp(-\tau q),\qquad \tau>0.
\]
Unlike the truncated-linear estimator it modifies TSLS for every value of $Q$, so Theorem \ref{thm: stabilization} does not apply. A separate argument shows that it also dominates TSLS when $\tau$ is large enough.

\begin{proposition}\label{prop: exponential}
Let $c=4\kappa(k-1)H^2/s_*$. If $1+2\tau>2\kappa(k-2)+c$, then for all $\rho\in\R^k$ and $\pi\neq0$,
\[
R_{\rho,\pi}(\delta_0)-R_{\rho,\pi}(\delta_{f_e})\ge s_*\left\{1-\frac{c}{1+2\tau-2\kappa(k-2)}\right\}\E_\pi\left[\frac{e^{-\tau Q}}{Q}\right]>0.
\]
\end{proposition}

The improvement from the exponential estimator is again positive at every parameter value, and its lower bound does not depend on $\rho$ and depends on $\pi$ only through the noncentrality $\pi'\Spp^{-1}\pi$. The threshold for $\tau$ grows linearly with the number of instruments. Some linear growth is necessary when $W$ is proportional to $\Spp^{-1}$: Step 1 of the proof of Proposition \ref{prop: bound} shows that dominance then requires $1+2\tau>k-2$. For TSLS under homoskedasticity, where $\kappa=1$ and $H=0$, the sufficient condition in Proposition \ref{prop: exponential} reduces to $\tau>k-5/2$, while the necessary condition is $\tau>(k-3)/2$. The proof, in Appendix \ref{app: proofs}, writes the improvement as a quadratic function of the purged reduced-form mean $a$ and shows that this quadratic is positive for every $a$. The coefficients of the quadratic are controlled by expanding the distribution of the direction of $\hpi$ in a power series in the first-stage mean, and the condition on $\tau$ ensures that the quadratic term dominates the linear term.

The conditions in Theorem \ref{thm: stabilization} and Proposition \ref{prop: exponential} are sufficient, and conservative. Appendix \ref{app: certificates} gives sharper sufficient conditions for both estimators. These take the form of a finite set of inequalities on one-dimensional moments of the taper, together with a constant computed from $\Spp$ and $W$, and they imply dominance at every parameter value. We verify them in exact rational arithmetic, so the resulting tuning parameters are certified to deliver dominance rather than checked on a grid of parameter values. In the three-instrument designs of Section \ref{sec: simulations}, for example, the certified cutoffs $Q_*$ are roughly 6 to 13 times larger than $\bar q$.

The improvements of both estimators are easiest to describe when $\rho=0$ and $\pi=0$, where the target is undefined but all of the estimators in \eqref{eq: taper} have mean $m$. Since the length and direction of the standard normal vector $\Spp^{-1/2}\hpi$ are independent when $\pi=0$,
\begin{equation}\label{eq: null variance}
\E_{0,0}[(\delta_f-m)^2]=\mathrm{Var}_{0,0}(b(\hpi))+C_S\,\E\left[f(X)^2\right],\qquad X\sim\chi^2_{k-2},
\end{equation}
where $C_S=\E_{0,0}[\ell(\hpi)'S\ell(\hpi)]$. The estimators in \eqref{eq: taper} therefore reduce only the second component of the TSLS variance, and the fraction of $C_S$ they remove depends only on $k$ and the taper $f$. For the truncated-linear estimator, the fraction of $C_S$ removed is $F_{k-2}(Q_*)-k(k-2)F_{k+2}(Q_*)/Q_*^2$, where $F_j$ is the $\chi^2_j$ distribution function. For the exponential estimator it is $2(1+2\tau)^{-(k-2)/2}-(1+4\tau)^{-(k-2)/2}$.

How large can these improvements be? Our next result bounds the variance reduction available from any estimator of the form \eqref{eq: taper} that dominates TSLS. For $c>0$ let $f_c(q)=\min\{1,q/c\}$, let $X\sim\chi^2_{k-2}$, and let $c_k$ denote the unique $c>1$ that solves
\[
\E\left[(1-X/c)(1+X/c-2X)\,1\{X<c\}\right]=0.
\]

\begin{proposition}\label{prop: bound}
Suppose $W$ is proportional to $\Spp^{-1}$. If $f:[0,\infty)\to[0,1]$ is measurable and $R_{\rho,\pi}(\delta_f)\le R_{\rho,\pi}(\delta_0)$ for all $\rho\in\R^k$ and $\pi\neq0$, then
\[
\E_{0,0}[(\delta_0-m)^2]-\E_{0,0}[(\delta_f-m)^2]\le\psi_k\,\E_{0,0}[(\delta_0-m)^2],\qquad \psi_k=\E\left[1-f_{c_k}(X)^2\right].
\]
Equality holds if and only if $f=f_{c_k}$ Lebesgue-almost everywhere on $(0,\infty)$ and $b$ is constant on $\R^k\setminus\{0\}$.
\end{proposition}

The condition that $W$ be proportional to $\Spp^{-1}$ holds for TSLS under homoskedasticity. The proof first shows that any dominating taper must satisfy $\E[1-f(X)^2]\ge2\E[\{1-f(X)\}X]$, by considering parameter values with $\pi$ approaching zero and $a$ large. It then maximizes the variance reduction subject to this constraint, which is a concave program whose solution is the truncated-linear taper with cutoff $c_k$. The bound $\psi_k$ equals 0.89 for $k=3$, 0.52 for $k=4$, 0.25 for $k=5$, 0.014 for $k=8$, and 0.0015 for $k=10$, and is below $10^{-9}$ for $k\ge21$, since $\psi_{21}\approx3.6\times10^{-10}$ and $\psi_k$ is nonincreasing in $k$. Proposition \ref{prop: bound} thus shows that, when $W$ is proportional to $\Spp^{-1}$, estimators of the form \eqref{eq: taper} that dominate TSLS can deliver large variance reductions at $\rho=\pi=0$ only when the number of instruments is small. It also shows that among such estimators the truncated-linear shape is the only candidate to attain the bound. In the designs of Section \ref{sec: simulations} with $k\ge21$, which nearly satisfy this condition ($\kappa\le1.016$), the certified truncated-linear cutoffs are within 1\% of $c_k$.

Neither estimator dominates the other in general. Appendix \ref{app: crossing} gives a homoskedastic design with three instruments, and tuning parameters satisfying the conditions of Theorem \ref{thm: stabilization} and Proposition \ref{prop: exponential}, for which each estimator has strictly lower risk than the other at some parameter values. The truncated-linear estimator leaves TSLS unchanged unless $Q$ is small, and is the only shape that can attain the bound in Proposition \ref{prop: bound}. The exponential estimator has less conservative explicit tuning: in Table \ref{tab: simulations} its explicit version removes 26\% of the variance at $\rho=\pi=0$ on average, against 12\% for the explicit truncated-linear estimator. In the two specifications in Section \ref{sec: simulations} where improvements away from zero are non-negligible, it also improves more than the truncated-linear estimator with the same type of tuning. We make no claim that either estimator is admissible or optimal.

\section{Simulations Calibrated to Published Specifications}\label{sec: simulations}

The results so far establish that TSLS can be improved at every parameter value by estimators that modify it only, or mainly, when the observed first stage is close to zero. To gauge their size in practice, we evaluate the four estimators from Section \ref{sec: dominators} in designs calibrated to published IV specifications. These are the truncated-linear and exponential estimators, each with explicit and with certified tuning.

We use the specifications from the American Economic Review between 2014 and 2018 collected by \citet{AndrewsStockSun2019} that have at least three instruments, so that TSLS has finite risk. This yields 70 specifications. Thirty come from \citet{StephensYang2014} and have three instruments. The remaining forty come from \citet{Young2014} and have 21, 28, or 59 instruments. Two papers is a narrow base, and we view the results as illustrative. For each specification we set $\Sigma$ equal to the estimated covariance matrix of the reduced-form and first-stage coefficients, and set $W$ to the TSLS weight. The four estimators are tuned once for each specification and then held fixed. For the explicit tunings we set $Q_*=\bar q$ and set $\tau$ to $1+10^{-4}$ times the threshold $\kappa(k-2)+c/2-1/2$ in Proposition \ref{prop: exponential}, using the certified bounds on $\kappa$, $s_*$, and $H$ described in Appendix \ref{app: certificates}.

We report three comparisons, each as the percentage reduction relative to TSLS. The first is the variance at $\rho=0$ and $\pi=0$, computed from \eqref{eq: null variance}. The second is the mean squared error at the estimated coefficients, $(\rho,\pi)=(\hrho,\hpi)$, with the target recomputed at those values. The third averages the mean squared error over the posterior for $(\rho,\pi)$ under a flat prior, $N((\hrho',\hpi')',\Sigma)$. We report the ratio of posterior-average risks, rather than the average of ratios, with each draw's target computed at that draw. Appendix \ref{app: numerics} describes the numerical integration, including the importance sampling we use for the posterior averages.

\begin{table}[t]
\centering
\begin{tabular}{lccc}
 & Variance at & MSE at & Posterior-average\\
 & $\rho=\pi=0$ & estimates & MSE\\
\hline\hline
Truncated-linear, explicit & 12.36 (14.74) & 0.00 (0.02) & 0.10 (0.64)\\
 & [36.34] & [0.17] & [5.08]\\
Truncated-linear, certified & 32.19 (38.16) & 0.03 (0.20) & 0.37 (2.35)\\
 & [86.52] & [1.64] & [18.29]\\
Exponential, explicit & 26.07 (31.19) & 0.03 (0.20) & 0.32 (2.09)\\
 & [77.82] & [1.65] & [16.45]\\
Exponential, certified & 31.17 (36.89) & 0.08 (0.48) & 0.46 (2.87)\\
 & [85.80] & [3.71] & [21.91]\\
\hline
\end{tabular}
\caption{Percentage reduction relative to TSLS across the 70 specifications with at least three instruments in the \citet{AndrewsStockSun2019} sample. Entries are the mean, with the standard deviation across specifications in parentheses and the maximum in brackets. The 40 specifications with 21 or more instruments contribute essentially nothing to any column.\label{tab: simulations}}
\end{table}

Table \ref{tab: simulations} reports the results. Each column averages over two very different groups of specifications. When $\rho=0$ and $\pi=0$, the certified estimators reduce variance by 75\% and 73\% on average in the 30 three-instrument specifications, and by up to 87\%. In the 40 specifications with 21 or more instruments the reductions are below $10^{-7}$ percent, in line with $\psi_k<10^{-9}$ for $k\ge21$. At the estimated coefficients the improvements are below 0.1\% on average and at most 3.7\%. Averaging over the posterior yields somewhat larger improvements, below 0.5\% on average and up to 22\%. These are driven almost entirely by two specifications, which reach 10\% and 22\% for the certified exponential estimator. All other specifications show improvements below 0.1\%.

Two features of these designs explain the pattern away from $\rho=\pi=0$. First, the estimators modify TSLS appreciably only when the Wald statistic $Q$ is small in absolute terms: the truncated-linear estimator only when $Q\le Q_*$, and the exponential estimator by a factor $e^{-\tau Q}$. For large $k$ this is much more restrictive than a small first-stage $F$ statistic $Q/k$. The specifications from \citet{Young2014} have weak first stages by the rule of thumb of \citet{StaigerStock1997}, with first-stage $F$ statistics between 1.4 and 6.8 in the data. Since $k$ is large, however, the observed Wald statistics range from about 50 to 400, and these are the noncentralities of $Q$ at the estimated coefficients. Values of $Q$ small enough for the estimators to act then essentially never arise. Second, the observed Wald statistics in the three-instrument specifications from \citet{StephensYang2014} range from 19 to 282. The two specifications with non-negligible improvements have the two smallest, 19 and 25, at which $Q\le Q_*$ has probability about $2\times10^{-3}$ and $3\times10^{-4}$ under the certified truncated-linear tuning. For the other 28, whose Wald statistics exceed 58, this probability is below $10^{-9}$. The ranking of the estimators also differs across columns. The certified truncated-linear estimator has the lowest variance at $\rho=\pi=0$ in most three-instrument specifications, while the certified exponential estimator has the largest improvements in the two specifications where the improvements away from zero are non-negligible.

\section{Conclusion}\label{sec: conclusion}

We show that TSLS is inadmissible under squared error loss whenever there are at least three instruments, for any covariance and weight matrix, and whether or not treatment effects are constant. Simple estimators that stabilize TSLS when the observed first stage is close to zero improve on it at every parameter value. In designs calibrated to published specifications, however, the resulting improvements in mean squared error are small except in the two specifications with the smallest first-stage Wald statistics.

\appendix

\section{Proofs}\label{app: proofs}

\paragraph{Notation.} Throughout, $k\ge3$. Write $Z=\Spp^{-1/2}\hpi\sim N(\eta,I_k)$ with $\eta=\Spp^{-1/2}\pi$, and $Y=\Spp^{-1/2}\ha\sim N(\xi,\Omega)$ with $\xi=\Spp^{-1/2}a$ and $\Omega=\Spp^{-1/2}S\Spp^{-1/2}$, so that $Z$ and $Y$ are independent. Let $K=\Spp^{1/2}W\Spp^{1/2}$, $L(z)=Kz/z'Kz$, and $\tilde b(z)=b(\Spp^{1/2}z)$. Then $Q=|Z|^2$, $\ell(\hpi)'\ha=L(Z)'Y$, $\delta_0=\tilde b(Z)+L(Z)'Y$, and $\theta=\tilde b(\eta)+L(\eta)'\xi$. The function $\tilde b$ is even and takes values in $[b_-,b_+]$, while $L$ is odd and satisfies $L(rz)=L(z)/r$ for $r>0$, $n'L(n)=1$, and $|n|^{-1}\le|L(n)|\le\kappa|n|^{-1}$. Since $\E|Z|^{-2}<\infty$ for $k\ge3$, every expectation below is finite. For $\eta=re$ with $|e|=1$, write $\theta-\tilde b(\eta)=L(\eta)'\xi=\eta'K\xi/\eta'K\eta$, so that
\begin{equation}\label{eq: projection identity}
e'K\xi=\{\theta-\tilde b(\eta)\}\,|\eta|\,e'Ke.
\end{equation}

\paragraph{Proof of Lemma \ref{lem: replacement}.} Write $Z=rn$ with $r=\sqrt Q$ and $|n|=1$, and let $f(n)=L(n)'\xi$ and $s(n)=L(n)'\Omega L(n)$. Since $|L(n)|\ge1$, $s(n)\ge s_*$. Conditional on $Z=rn$, $\delta_0-\theta=\tilde b(n)-\theta+L(n)'Y/r$, so
\begin{equation}\label{eq: conditional change}
\E\left[(\tilde b(Z)-\theta)^2-(\delta_0-\theta)^2\mid Z=rn\right]=-\frac{2}{r}\{\tilde b(n)-\theta\}f(n)-\frac{f(n)^2}{r^2}-\frac{s(n)}{r^2}.
\end{equation}
Let $b_c=(b_++b_-)/2$, $t=\theta-b_c$, and $T=\sqrt{H^2+\kappa s_*}$, and note that $\bar q=s_*/\{2(H+T)^2\}=(T-H)/\{2\kappa(T+H)\}$ since $T^2-H^2=\kappa s_*$. Fix $r^2\le\bar q$.

Suppose first that $|t|\le T$. Then $|\tilde b(n)-\theta|\le H+T$, and the right side of \eqref{eq: conditional change} equals $(\tilde b(n)-\theta)^2-(\tilde b(n)-\theta+f(n)/r)^2-s(n)/r^2\le(H+T)^2-s_*/r^2\le-s_*/(2r^2)$ for every $n$.

Suppose instead that $|t|>T$. Conditional on $|Z|=r$, the direction $n$ has density proportional to $\exp(r\eta'n)$ with respect to the uniform measure $\sigma$ on the unit sphere. Since $\tilde b$ is even and $f$ is odd, pairing $n$ with $-n$ gives
\begin{align*}
\E\left[\eqref{eq: conditional change}\mid Q=r^2\right]=\frac{1}{Z_r}\int\Big[&-\frac{2}{r}\{\tilde b(n)-\theta\}f(n)\sinh(r\eta'n)\\
&-\frac{f(n)^2+s(n)}{r^2}\cosh(r\eta'n)\Big]\,d\sigma(n),
\end{align*}
where $Z_r=\int\cosh(r\eta'n)\,d\sigma(n)$. Let
\[
A_r=\int f(n)^2\cosh(r\eta'n)\,d\sigma(n),\qquad J_r=\int(\eta'n)^2\cosh(r\eta'n)\,d\sigma(n).
\]
Since $|\sinh x|\le|x|\cosh x$ and $|\tilde b(n)-\theta|\le|t|+H$, the Cauchy--Schwarz inequality gives
\[
\left|\frac{2}{r}\int\{\tilde b(n)-\theta\}f(n)\sinh(r\eta'n)\,d\sigma(n)\right|\le2(|t|+H)\sqrt{A_rJ_r}.
\]
To bound $J_r$ by $A_r$, let $e=\eta/|\eta|$ and $N_r=\int nn'\cosh(r\eta'n)\,d\sigma(n)$. By rotational symmetry about $e$, $N_r=\lambda_\parallel ee'+\lambda_\perp(I-ee')$ with $\lambda_\perp\ge0$, so $J_r=|\eta|^2\lambda_\parallel$. Since $n'Kn\le\lambda_{\max}(K)$, \eqref{eq: projection identity} gives
\[
A_r\ge\frac{\xi'KN_rK\xi}{\lambda_{\max}(K)^2}\ge\frac{\lambda_\parallel(e'K\xi)^2}{\lambda_{\max}(K)^2}=\frac{J_r\{\theta-\tilde b(\eta)\}^2(e'Ke)^2}{\lambda_{\max}(K)^2}\ge\frac{(|t|-H)^2}{\kappa^2}J_r,
\]
where the last step uses $|\theta-\tilde b(\eta)|\ge|t|-H>0$ and $e'Ke\ge\lambda_{\min}(K)$. Hence $\sqrt{A_rJ_r}\le\kappa A_r/(|t|-H)$, and
\[
\E\left[\eqref{eq: conditional change}\mid Q=r^2\right]\le-\frac{s_*}{r^2}+\frac{A_r}{Z_r}\left\{-\frac{1}{r^2}+\frac{2\kappa(|t|+H)}{|t|-H}\right\}.
\]
Since $(|t|-H)/(|t|+H)$ is nondecreasing in $|t|$, it is at least $(T-H)/(T+H)=2\kappa\bar q$, so the term in braces is nonpositive when $r^2\le\bar q$. In both cases the conditional expectation is at most $-s_*/(2r^2)$. $\square$

\paragraph{Proof of Theorem \ref{thm: stabilization}.} Write $w=w(Q)$. Since $\delta_w-\theta=(1-w)(\delta_0-\theta)+w\{\tilde b(Z)-\theta\}$ and squared error is convex, $(\delta_0-\theta)^2-(\delta_w-\theta)^2\ge w[(\delta_0-\theta)^2-\{\tilde b(Z)-\theta\}^2]$. Taking expectations conditional on $Q$ and applying Lemma \ref{lem: replacement} on the event $Q\le\bar q$, outside of which $w=0$, gives $\E[(\delta_0-\theta)^2-(\delta_w-\theta)^2\mid Q]\ge s_*w(Q)/(2Q)$. Taking expectations proves the inequality. Since $Q$ has a positive density on $(0,\infty)$ and $w$ is positive on a set of positive measure, $\E_\pi[w(Q)/Q]>0$. $\square$

\paragraph{Proof of Corollary \ref{cor: inadmissible}.} The function $w(q)=(1-q/\bar q)_+$ satisfies the conditions of Theorem \ref{thm: stabilization}. The constant-effects model is the subset of the parameter space with $\rho=\beta\pi$, on which $\theta=\beta$. $\square$

The proofs of Theorems \ref{thm: calibration} and \ref{thm: admissible} use the following continuity property. We work with $(\ha,\hpi)\sim N(\mu,\Lambda)$, where $\mu=(a',\pi')'$ and $\Lambda=\mathrm{diag}(S,\Spp)$, and note that $\rho=a$ when $\pi=0$.

\begin{lemma}\label{lem: continuity}
Suppose $\E_{a,\pi}[\delta^2]<\infty$ for all $a$ and all $\pi\neq0$. Then $\E_{a,\pi}[\delta^2]<\infty$ for all $(a,\pi)\in\R^{2k}$, and $\E_{a,\pi}[\delta]$ and $\E_{a,\pi}[\delta^2]$ are continuous on $\R^{2k}$.
\end{lemma}

\begin{proof}
The density of $N(\mu,\Lambda)$ is $\phi_0(x)\exp(\tau'x-\tau'\Lambda\tau/2)$, where $\tau=\Lambda^{-1}\mu$. Fix a compact set of values of $\tau$, and enclose it in a box whose vertices $\tau_1,\ldots,\tau_J$ all have nonzero last $k$ coordinates. Since $\tau\mapsto\exp(\tau'x)$ is convex, its maximum over the box is attained at a vertex, so $\exp(\tau'x)\le\sum_j\exp(\tau_j'x)$. Consequently, uniformly over the compact set, the density at $\mu$ is bounded by a constant times $\sum_j\phi_{\mu_j}(x)$, where $\mu_j=\Lambda\tau_j$ has a nonzero first-stage component. The right side integrates $1+|\delta|+\delta^2$, and the conclusion follows by dominated convergence.
\end{proof}

\paragraph{Proof of Theorem \ref{thm: calibration}.} Since $R_{\rho,\pi}(\delta)\le R_{\rho,\pi}(\delta_0)<\infty$, $\delta$ has a finite second moment at every parameter value with $\pi\neq0$, so Lemma \ref{lem: continuity} applies to both $\delta$ and $\delta_0$. Fix $a\in\R^k$ and $t\in\R$, choose $v\neq0$ with $v'Wa=0$, and for $\varepsilon>0$ let $\pi_\varepsilon=\varepsilon v$ and $a_\varepsilon=a+\varepsilon(tI-\Gamma)v$. By \eqref{eq: decomposition},
\[
\theta=b(v)+\frac{v'Wa+\varepsilon v'W(tI-\Gamma)v}{\varepsilon v'Wv}=t
\]
along this sequence. Letting $\varepsilon\downarrow0$ in $R(\delta)\le R(\delta_0)$ and applying Lemma \ref{lem: continuity} gives $\E_{a,0}[(\delta-t)^2]\le\E_{a,0}[(\delta_0-t)^2]$ for every $t\in\R$. The difference between the two sides is affine in $t$, so $\E_{a,0}[\delta]=\E_{a,0}[\delta_0]$ and $\E_{a,0}[\delta^2]\le\E_{a,0}[\delta_0^2]$, which proves the second part. Since $b$ is even, $\ell$ is odd, and $\E|\ell(\hpi)|<\infty$, $\E_{a,0}[\delta_0]=m$. Let $r(u)=\E_{V\sim N(0,\Spp)}[\delta(u,V)]-m$. By Jensen's inequality $r$ is square-integrable under $N(a,S)$ for every $a$, and we have shown that $\E_{\ha\sim N(a,S)}[r(\ha)]=0$ for every $a\in\R^k$. Completeness of the normal location family \citep[e.g.][Theorem 4.3.1]{LehmannRomano2005} then implies $r(u)=0$ for almost every $u$. $\square$

\paragraph{Proof of Theorem \ref{thm: admissible}.} Every $\delta\in\mathcal D_{\tilde\delta}$ satisfies $R(\delta)\le R(\tilde\delta)\le R(\delta_0)$, so by Theorem \ref{thm: calibration} it is square-integrable under $P_{0,0}$, the distribution with $\rho=0$ and $\pi=0$. We view $\mathcal D_{\tilde\delta}$ as a subset of $L^2(P_{0,0})$, identifying estimators that agree $P_{0,0}$-almost surely. All of the normal distributions in \eqref{eq: model} are mutually absolutely continuous, so such estimators have the same risk at every parameter value. The set $\mathcal D_{\tilde\delta}$ is nonempty since it contains $\tilde\delta$, and convex since squared error is convex. It is closed: if $\delta_n\to \delta$ in $L^2(P_{0,0})$, a subsequence converges $P_{0,0}$-almost surely, hence almost surely under every distribution in the model, and Fatou's lemma gives $R(\delta)\le\liminf R(\delta_n)\le R(\tilde\delta)$. The objective $\E_{0,0}[(\delta-m)^2]$ is the squared distance from $\delta$ to the constant $m$, so it has a unique minimizer $\delta^*$ over the closed convex set $\mathcal D_{\tilde\delta}$.

Suppose $\delta'$ dominates $\delta^*$. Then $R(\delta')\le R(\delta^*)\le R(\tilde\delta)$, so $\delta'\in\mathcal D_{\tilde\delta}$. For any $v\neq0$, along $\pi_\varepsilon=\varepsilon v$ and $a_\varepsilon=\varepsilon(mI-\Gamma)v$ the target equals $m$ by \eqref{eq: decomposition}, and both $\rho_\varepsilon$ and $\pi_\varepsilon$ tend to zero. Letting $\varepsilon\downarrow0$ in $R(\delta')\le R(\delta^*)$ and applying Lemma \ref{lem: continuity} gives $\E_{0,0}[(\delta'-m)^2]\le\E_{0,0}[(\delta^*-m)^2]$. By uniqueness of the minimizer $\delta'=\delta^*$ almost surely, so the two have the same risk everywhere, contradicting the assumption that $\delta'$ dominates $\delta^*$. Hence $\delta^*$ is admissible. If $\tilde\delta=\delta_w$, then $R(\delta^*)\le R(\delta_w)<R(\delta_0)$ at every parameter value. $\square$

\paragraph{Gain identity.} The proofs of Propositions \ref{prop: exponential} and \ref{prop: bound} use an expression for the improvement of $\delta_f$ over TSLS. Let $g=1-f$ and $h=1-f^2=2g-g^2$, both evaluated at $Q$. Conditional on $Z$, $(\delta_0-\theta)^2-(\delta_f-\theta)^2=h\,(L'Y)^2+2g\{\tilde b(Z)-\theta\}L'Y$, where $L=L(Z)$. Taking expectations over $Y$, substituting $\theta=\tilde b(\eta)+L(\eta)'\xi$, and then taking expectations over $Z$ gives
\begin{equation}\label{eq: gain identity}
R(\delta_0)-R(\delta_f)=\xi'A\xi+2\xi'z+c_0,
\end{equation}
where $A=T_h-m_gL(\eta)'-L(\eta)m_g'$, with $T_h=\E[hLL']$, $m_g=\E[gL]$, $z=\E[g\{\tilde b(Z)-\tilde b(\eta)\}L]$, and $c_0=\E[hL'\Omega L]$. Here expectations are over $Z\sim N(\eta,I)$. Two facts are used repeatedly. First, the density of $\chi^2_k$ divided by its argument is the density of $\chi^2_{k-2}$ divided by $k-2$, so $\E_{\chi^2_k}[u(X)/X]=\E_{\chi^2_{k-2}}[u(X)]/(k-2)$. Second, under $N(0,I_k)$ the length $|Z|$ and the direction $Z/|Z|$ are independent.

\paragraph{Proof of Proposition \ref{prop: exponential}.} Let $g(q)=e^{-\tau q}$, $\alpha=1+2\tau$, $d=k-2$, and $J=\E[g/Q]$. Let $n$ be uniform on the unit sphere. Write $\eta=re$ with $|e|=1$, and define $M_j=\E_n[L(n)L(n)'(e'n)^{2j}]$ and $v_j=\E_n[L(n)(e'n)^{2j+1}]$ for $j\ge0$.

We first bound the quadratic term. Matrix Cauchy--Schwarz gives $v_j'M_j^{-1}v_j\le\omega_j=\E_n[(e'n)^{2j+2}]$. Since $n'Kn\le\lambda_{\max}(K)$, $M_j\succeq\lambda_{\max}(K)^{-2}KN_jK$ with $N_j=\E_n[nn'(e'n)^{2j}]$, and $N_je=\omega_je$ by symmetry, so $L(e)'M_j^{-1}L(e)\le\kappa^2/\omega_j$. Since the largest eigenvalue of $xy'+yx'$ is at most $2|x||y|$, these bounds give
\begin{equation}\label{eq: angular}
v_jL(e)'+L(e)v_j'\preceq2\kappa M_j.
\end{equation}
Expanding $\exp(r e'Z)$ in the density of $N(re,I)$ relative to $N(0,I)$, and keeping the even and odd terms respectively, gives the absolutely convergent series
\begin{align*}
T_h&=e^{-r^2/2}\sum_{j\ge0}\frac{r^{2j}}{(2j)!}\E[h(X)X^{j-1}]M_j,\\
m_gL(\eta)'+L(\eta)m_g'&=e^{-r^2/2}\sum_{j\ge0}\frac{r^{2j}}{(2j)!}\frac{\E[g(X)X^j]}{2j+1}\\
&\qquad\times\{v_jL(e)'+L(e)v_j'\},
\end{align*}
where $X\sim\chi^2_k$. Gamma integration gives $\E[g(X)X^j]=\{(k+2j-2)/\alpha\}\E[g(X)X^{j-1}]$, and $\E[h(X)X^{j-1}]\ge\E[g(X)X^{j-1}]$ since $h\ge g$. Since $(k+2j-2)/(2j+1)\le d$ for $k\ge3$, \eqref{eq: angular} yields $A\succeq(1-2\kappa d/\alpha)T_h$, which is positive definite under the condition of the proposition.

We next bound the linear term. Let $\Delta=\tilde b(Z)-\tilde b(\eta)$. In the coordinates $x=K^{1/2}z$, $\tilde b$ is a Rayleigh quotient of a symmetric matrix with eigenvalues in $[b_-,b_+]$, so the difference between its values at two unit vectors is at most $2H$ times the sine of the angle between them. Since $\sin\angle(K^{1/2}z,K^{1/2}\eta)\le\kappa^{1/2}\sin\angle(z,\eta)$, we obtain $\Delta^2\le4\kappa H^2\sin^2\angle(Z,\eta)$. Let $Z_i$ be a coordinate of $Z$ orthogonal to $\eta$. Gaussian integration by parts applied to $g(Q)Z_i/Q$ gives $\E[(g-2g')Z_i^2/Q]+2\E[gZ_i^2/Q^2]=\E[g/Q]$. The boundary term near the origin vanishes since $k\ge3$. Since $g-2g'=\alpha g$, summing over the $k-1$ orthogonal coordinates gives $\E[g\sin^2\angle(Z,\eta)]\le(k-1)J/\alpha$. By matrix Cauchy--Schwarz and $g^2/h=g/(2-g)\le g$,
\[
zz'\preceq T_h\E\left[\frac{g^2}{h}\Delta^2\right]\preceq\frac{4\kappa(k-1)H^2}{\alpha}J\,T_h=\frac{cs_*}{\alpha}J\,T_h.
\]
Combining with the bound on $A$, $z'A^{-1}z\le cs_*J/(\alpha-2\kappa d)$. Finally, $c_0\ge s_*\E[h/Q]\ge s_*J$ since $L'\Omega L\ge s_*/Q$ and $h\ge g$. Completing the square in \eqref{eq: gain identity},
\[
R(\delta_0)-R(\delta_{f_e})\ge c_0-z'A^{-1}z\ge s_*J\left\{1-\frac{c}{\alpha-2\kappa d}\right\}>0.\qquad\square
\]

\paragraph{Proof of Proposition \ref{prop: bound}.} Since $W$ is proportional to $\Spp^{-1}$, $K$ is proportional to the identity and $L(z)=z/|z|^2$. Let $X\sim\chi^2_{k-2}$, $g=1-f$, and $h=1-f^2$.

Step 1 (a necessary condition). Since $\xi$ is unrestricted, dominance and \eqref{eq: gain identity} imply that $A$ is positive semidefinite at every $\eta\neq0$: otherwise $R(\delta_0)-R(\delta_f)$ would be negative for $\xi$ a large multiple of a direction with $\xi'A\xi<0$. Take $\eta=re_1$, so that $L(\eta)_1=1/r$ and $e_1'Ae_1=\E_{re_1}[hZ_1^2/Q^2]-(2/r)\E_{re_1}[gZ_1/Q]$. Write $\E_{re_1}[u(Z)]=e^{-r^2/2}\E_0[u(Z)e^{rZ_1}]$. In the first term the integrand is bounded by $e^{|Z|}/|Z|^2$, which is integrable. In the second, $gZ_1/Q$ is odd, so $(2/r)\E_0[gZ_1e^{rZ_1}/Q]=2\E_0[gZ_1\sinh(rZ_1)/(rQ)]$. Since $|\sinh(rx)/r|\le|x|e^{|x|}$ for $0<r\le1$, the integrand of the last expectation is bounded by $e^{|Z|}$. By dominated convergence and independence of the length and direction of $Z$ under $N(0,I)$,
\[
\lim_{r\downarrow0}e_1'Ae_1=\frac{1}{k}\left\{\E_{\chi^2_k}\left[\frac{h(Q)}{Q}\right]-2\E_{\chi^2_k}[g(Q)]\right\}=\frac{1}{k(k-2)}\left\{\E[h(X)]-2\E[g(X)X]\right\}\ge0.
\]

Step 2 (the variance reduction). Under $\rho=0$ and $\pi=0$, $Y\sim N(0,\Omega)$ is independent of $Z\sim N(0,I)$, so by \eqref{eq: null variance}, $\E_{0,0}[(\delta_0-m)^2]=\mathrm{Var}_{0,0}(b(\hpi))+C_S$ and $\E_{0,0}[(\delta_0-m)^2]-\E_{0,0}[(\delta_f-m)^2]=C_S\E[h(X)]$. Here $C_S>0$, since $S$ is positive definite and $\ell(\hpi)\neq0$ almost surely. It therefore suffices to show that $\E[h(X)]\le\psi_k$ for every $g$ satisfying Step 1, with equality if and only if $f=f_{c_k}$ almost everywhere. The variance reduction is then $C_S\E[h(X)]\le C_S\psi_k\le\psi_k\E_{0,0}[(\delta_0-m)^2]$. Since $\psi_k>0$, both inequalities hold with equality if and only if $f=f_{c_k}$ almost everywhere and $\mathrm{Var}_{0,0}(b(\hpi))=0$. Since $b$ is continuous away from zero and $\hpi$ has a positive density, the variance is zero if and only if $b$ is constant on $\R^k\setminus\{0\}$.

Step 3 (duality). Let $g$ satisfy Step 1 and let $\mu>0$. Then $\E[h(X)]\le\E[(1+\mu)h(X)-2\mu g(X)X]$. For each $x\ge0$ the function $\gamma\mapsto(1+\mu)(2\gamma-\gamma^2)-2\mu\gamma x$ is strictly concave, with unique maximizer over $[0,1]$ equal to $(1-x/c)_+$, where $c=(1+\mu)/\mu>1$. Let $g_c=(1-x/c)_+$ and $h_c=2g_c-g_c^2$, and let $\Phi(c)=\E[h_c(X)-2g_c(X)X]$. Then $\E[h(X)]\le\E[h_c(X)]+\mu\Phi(c)$. If $\Phi(c)=0$, then $g_c$ itself satisfies Step 1 and the bound equals $\E[h_c(X)]$. Equality for a feasible $g$ requires $g$ to attain the pointwise maximum almost everywhere, so $g=g_c$ almost everywhere.

Step 4 (the root). The function $\Phi$ is continuous by dominated convergence. For $c\le1$ and $x<c$, $1+x/c-2x\ge1-x>0$, so $\Phi>0$ on $(0,1]$. As $c\to\infty$, $\Phi(c)\to\E[1-2X]=1-2(k-2)<0$. Hence $\Phi$ has a root $c_k>1$. If $c$ and $c'$ were two roots, Step 3 would give $g_{c'}=g_c$ almost everywhere, and since $X$ has a positive density on $(0,\infty)$, $c'=c$. Writing $1-f_{c_k}^2=h_{c_k}$, Steps 2 to 4 prove the proposition. Finally, $(1+\mu)h_c(x)-2\mu g_c(x)x=(1+\mu)g_c(x)^2$ for every $x\ge0$, and $1+\mu=c/(c-1)$. Step 3 applied to $g=g_{c_k}$ therefore gives $\psi_k\le\{c/(c-1)\}\E[g_c(X)^2]$ for every $c>1$, with equality at $c=c_k$ because $\Phi(c_k)=0$. Since $g_c(x)^2$ is nonincreasing in $x$ and $\chi^2_{k-2}$ is stochastically increasing in $k$, $\psi_k$ is nonincreasing in $k$. $\square$

\section{Sharper Certificates}\label{app: certificates}

This appendix gives the sufficient conditions behind the certified tuning parameters in Section \ref{sec: simulations}. We use the notation of Appendix \ref{app: proofs}. Let $X\sim\chi^2_{k-2}$, and for a taper $f$ let $g=1-f$, $h=1-f^2$, $U_m=\E[g(X)X^m]$, and $H_m=\E[h(X)X^m]$ for $m\ge0$. With $n$ uniform on the unit sphere, let $M_0=\E_n[L(n)L(n)']$ and $N=\E_n[L(n)n']$, and let $C_0$ be any constant such that
\begin{equation}\label{eq: central angular}
Ne\,L(e)'+L(e)\,e'N'\preceq C_0M_0\qquad\text{for all unit vectors }e.
\end{equation}
By \eqref{eq: angular} with $j=0$, $C_0=2\kappa$ is always valid, but smaller values can often be certified, as discussed below. Recall $c=4\kappa(k-1)H^2/s_*$.

\begin{proposition}\label{prop: certificate}
Let $g:[0,\infty)\to[0,1]$ be nonincreasing, absolutely continuous, not almost everywhere zero, and zero on $(Q_*,\infty)$. Suppose there exist $\zeta>0$, $\bar r>0$, and $\nu\in(0,1)$ such that
\begin{enumerate}
\item $-2g'(q)\ge(\zeta-1)g(q)$ for almost every $q$;
\item $H_0\ge\bar rU_0$;
\item $H_0-\nu U_0\ge C_0U_1$;
\item $(2m+1)(H_m-\nu U_m)\ge2\kappa U_{m+1}$ for $1\le m<\bar M$, where $\bar M=\max\{1,\lceil\kappa Q_*/(1-\nu)-1/2\rceil\}$; and
\item $\bar r\nu>c/\zeta$.
\end{enumerate}
Then $\delta_f$ strictly dominates TSLS.
\end{proposition}

\begin{proof}
We follow the proof of Proposition \ref{prop: exponential}, with sharper bounds at each step. Converting the series expansions there to moments of $X$, the coefficient of $M_j$ in $T_h$ is proportional to $H_j$, and that of $v_jL(e)'+L(e)v_j'$ in $m_gL(\eta)'+L(\eta)m_g'$ is proportional to $U_{j+1}/(2j+1)$, with the same positive proportionality factor. Using \eqref{eq: central angular} for $j=0$ and \eqref{eq: angular} for $j\ge1$, conditions 3 and 4 imply that each coefficient of $A$ dominates $\nu$ times the corresponding coefficient of $T_g=\E[gLL']$. For $m\ge\bar M$ this holds automatically, since $H_m\ge U_m$ and $U_{m+1}\le Q_*U_m$. Hence $A\succeq\nu T_g$. Since $2-g$ is nondecreasing, tilting the distribution proportional to $g(X)$ by $X^m$ increases the mean of $2-g$, so $H_m/U_m\ge H_0/U_0\ge\bar r$ for all $m$. Applying this in the series gives $T_h\succeq\bar rT_g$ and $c_0\ge s_*\bar rJ$, where $J=\E[g/Q]$. Under condition 1, $g-2g'\ge\zeta g$, so the integration by parts argument gives $\E[g\sin^2\angle(Z,\eta)]\le(k-1)J/\zeta$. Matrix Cauchy--Schwarz with weight $g$ gives $zz'\preceq(cs_*J/\zeta)T_g$, so $z'A^{-1}z\le cs_*J/(\zeta\nu)$. Completing the square in \eqref{eq: gain identity} gives $R(\delta_0)-R(\delta_f)\ge s_*J(\bar r-c/(\zeta\nu))>0$. Since $g$ vanishes at $Q_*$ and is absolutely continuous, the integration by parts has no boundary term at the cutoff.
\end{proof}

\paragraph{The truncated-linear estimator.} For $g(q)=(1-q/Q_*)_+$ condition 1 holds with $\zeta=1+2/Q_*$. With $x=q/Q_*$ and $I_\beta(Q_*)=\int_0^1x^{\beta-1}(1-x)e^{-Q_*x/2}dx$, we have $H_m/U_m=1+e_m$ and $U_{m+1}/U_m=Q_*e_m$, where $e_m=I_{(k-2)/2+m+1}(Q_*)/I_{(k-2)/2+m}(Q_*)$. Conditions 2 to 5 therefore reduce to finitely many inequalities between one-dimensional integrals, which we bound above and below by rational numbers using alternating Taylor expansions of $e^{-y}$. We locate numerically the supremum of the cutoffs passing these conditions, reduce it by a relative reserve of $10^{-5}$, and verify all conditions at the reduced value in exact arithmetic.

\paragraph{The exponential estimator.} For $g(q)=e^{-\tau q}$, condition 1 holds with $\zeta=1+2\tau$. Gamma integration gives $H_m/U_m=2-v^{(k-2)/2+m}$ with $v=(1+2\tau)/(1+4\tau)$, and $U_{m+1}/U_m=(k-2+2m)/(1+2\tau)$. Since $H_m/U_m$ increases in $m$ and $(k-2+2m)/(2m+1)$ is nonincreasing for $k\ge3$, condition 4 for every $m\ge1$ follows from the case $m=1$. The argument of Proposition \ref{prop: certificate} then applies without the support restriction. With $d=k-2$ and $\alpha=1+2\tau$, so that $v=\alpha/(2\alpha-1)$, and with $\bar r=2-v^{d/2}$, it therefore suffices that some $\nu\in(0,1)$ satisfy
\[
\bar r-\nu\ge\frac{C_0d}{\alpha},\qquad 3\left(2-v^{d/2+1}-\nu\right)\ge\frac{2\kappa(d+2)}{\alpha},\qquad \bar r\nu>\frac{c}{\alpha}.
\]
We take the smallest $\tau$, up to a small reserve, that passes these conditions.

\paragraph{Certified constants.} The constants $\kappa$, $s_*$, and $H$ enter only through upper bounds on $\kappa$ and $H$ and a lower bound on $s_*$. We obtain these from rational matrix inequalities, $m_K\Spp^{-1}\preceq W\preceq M_K\Spp^{-1}$, $S\succeq\underline s\,\Spp$, and $b_-W\preceq(W\Gamma+\Gamma'W)/2\preceq b_+W$, which give $\kappa\le M_K/m_K$, $s_*\ge\underline s$, and $H\le(b_+-b_-)/2$. We check these inequalities by exact $LDL'$ factorization. For $C_0$, let $K=O\Lambda O'$ with $\Lambda=\mathrm{diag}(\lambda_1,\ldots,\lambda_k)$. For $V\sim N(0,I_k)$ and $s=V'\Lambda V$, define $D_{1j}=\E[V_j^2/s]$ and $D_{2j}=\E[V_j^2/s^2]$. Then \eqref{eq: central angular} holds whenever, for some $t>0$,
\[
C_0\ge\max_j\frac{(tD_{1j}+1)^2}{2t(k-2)D_{2j}\lambda_j}.
\]
This follows from the inequality $xy'+yx'\preceq(tx+y)(tx+y)'/(2t)$, after a congruence that diagonalizes the problem. The quantities $D_{1j}$ and $D_{2j}$ are one-dimensional integrals, which we enclose by interval Taylor integration. The eigenvalues are enclosed using a validated eigenbasis, and the final inequalities are checked in exact arithmetic. Across our designs the certified values of $C_0$ range from 2.0 to 2.7, while $2\kappa$ reaches about 9. As an independent check, we recomputed $\kappa$, $s_*$, $H$, and $C_0$ from the exact covariance and weight matrices for all 70 designs, and confirmed that all four tuned estimators pass their certificates in every design.

\section{Neither Estimator Dominates the Other}\label{app: crossing}

Let $k=3$, $\Sigma=I_6$, and $W=I_3$. Then $\Gamma=0$, $S=I_3$, $b\equiv0$, $\kappa=1$, $s_*=1$, $H=0$, and $\bar q=1/2$. Consider the truncated-linear estimator with $Q_*=1/2$, which dominates TSLS by Theorem \ref{thm: stabilization}. Consider also the exponential estimator with $\tau=3/5$, which dominates TSLS by Proposition \ref{prop: exponential}, since $1+2\tau=11/5>2$.

Let $a=\beta e_1$ and $\pi=re_1$. Since $b\equiv0$, $z=0$ in \eqref{eq: gain identity}, so the improvement over TSLS is $\beta^2e_1'Ae_1+c_0$. The limit computed in the proof of Proposition \ref{prop: bound} applies here, and the same dominated convergence argument gives $c_0\to\E_{\chi^2_3}[h(Q)/Q]$ as $r\downarrow0$. Hence, as $r\downarrow0$, the improvement of $\delta_f$ converges to
\[
G_f(\beta)=\mathcal H_f+\frac{\beta^2}{3}(\mathcal H_f-2\mathcal M_f),\qquad\mathcal H_f=\E_{\chi^2_3}\left[\frac{1-f(Q)^2}{Q}\right],\quad\mathcal M_f=\E_{\chi^2_3}[1-f(Q)].
\]
For $\beta\neq0$ both risks diverge as $r\downarrow0$, since the target $\beta/r$ diverges, but their difference converges.

For the exponential estimator, with $t=1+2\tau=11/5$ and $u=1+4\tau=17/5$, gamma integration gives $\mathcal H_{f_e}=2t^{-1/2}-u^{-1/2}$ and $\mathcal M_{f_e}=t^{-3/2}$. For the truncated-linear estimator, the substitution $Q=s^2/2$ gives
\[
\mathcal H_{f_c}=\pi^{-1/2}\int_0^1(1-s^4)e^{-s^2/4}ds,\qquad\mathcal H_{f_c}-2\mathcal M_{f_c}=\pi^{-1/2}\int_0^1(1-s^2)e^{-s^2/4}ds,
\]
where, here and in the bounds below, $\pi\approx3.14159$ denotes the mathematical constant rather than the first-stage coefficient. The following bounds use only rational arithmetic and the facts $3<\pi<3.15$, $1.4832<\sqrt{11/5}<1.4833$, and $1.8439<\sqrt{17/5}<1.8440$.
At $\beta=0$, $G_{f_e}(0)=2t^{-1/2}-u^{-1/2}>2/1.4833-1/1.8439>0.806$, while $G_{f_c}(0)\le\pi^{-1/2}\int_0^1(1-s^4)ds<0.8/\sqrt3<0.47$.

At $\beta=3$, $G_{f_e}(3)=8t^{-1/2}-4u^{-1/2}-6t^{-3/2}<8/1.4832-4/1.8440-6/(2.2\cdot1.4833)<1.386$. Since $4-3s^2-s^4\ge0$ on $[0,1]$ and $e^{-y}\ge1-y$, we have $G_{f_c}(3)=\pi^{-1/2}\int_0^1(4-3s^2-s^4)e^{-s^2/4}ds\ge\pi^{-1/2}\int_0^1(4-3s^2-s^4)(1-s^2/4)ds$. The right side equals $(4-4/3-1/20+1/28)/\sqrt\pi>2.652/\sqrt{3.15}>1.49$.
Hence for all sufficiently small $r>0$, the exponential estimator has strictly lower risk than the truncated-linear estimator at $(a,\pi)=(0,re_1)$, and strictly higher risk at $(a,\pi)=(3e_1,re_1)$. Numerically, the ordering already holds at $r=0.2$. There the improvements over TSLS are 0.793 for the exponential and 0.418 for the truncated-linear estimator when $\beta=0$, and 1.367 and 1.473 when $\beta=3$.

\section{Numerical Methods}\label{app: numerics}

\paragraph{Risk of TSLS and the exponential estimator.} Both risks reduce to one-dimensional integrals. Given $\hpi$, the reduced form enters quadratically and can be integrated exactly. The remaining expectations involve ratios of quadratic forms in $\hpi$, multiplied for the exponential estimator by $\exp(-\tau Q)$. Writing $1/s=\int_0^\infty e^{-xs}dx$ and $1/s^2=\int_0^\infty xe^{-xs}dx$ for $s=\hpi'W\hpi$, each such expectation is a one-dimensional integral of a Gaussian moment, which we evaluate by adaptive quadrature. We compute the improvement of the exponential estimator directly, rather than as a difference of risks, which preserves accuracy when the improvement is small.

\paragraph{Risk of the truncated-linear estimator.} After integrating out the reduced form, the improvement over TSLS is an integral over $\hpi$ restricted to the ball $Q\le Q_*$. In polar coordinates the radial integral is over $[0,\sqrt{Q_*}]$, and pairing antipodal directions replaces the exponential tilt by hyperbolic sines and cosines. With three instruments we use a product Gauss--Legendre rule, in the radius and in the cosine of the angle to $\Spp^{-1/2}\pi$, combined with a uniform azimuth grid. We compare a coarse grid with a fine one. With more instruments we combine the same radial rule with Monte Carlo over directions. The resulting MSE improvements, at the estimates and at the posterior draws, are below $10^{-21}$ as a fraction of TSLS risk. We also bound them analytically using $|L(z)|\le\kappa/|z|$ and the stochastic ordering of noncentral $\chi^2$ distributions.

\paragraph{Posterior averages.} With three instruments, the risk of TSLS generically grows like $1/|\pi|^2$ as $\pi\to0$, as does the squared target. Its posterior average is therefore finite, but its square is not integrable, so a plain Monte Carlo average over posterior draws has infinite variance. We instead draw the first-stage coefficients from a mixture of the posterior and a distribution whose density near the origin is proportional to $|\Spp^{-1/2}\pi|^{-2}$, and draw $a$ from its exact posterior. Reweighting by the ratio of densities yields an unbiased estimator with finite variance, and the delta method gives standard errors for the ratio of averages. Using 20,000 draws for each design, the standard error of each design's posterior-average percentage reduction is below 0.36 percentage points. A separate plain Monte Carlo run with 40,000 posterior draws agrees with the importance-sampling estimates to within 1.6 importance-sampling standard errors in the two specifications where the posterior improvements are non-negligible. For designs with more instruments, the squared risk is integrable and we use 400 posterior draws.

\paragraph{Validation.} Our TSLS risks and certified exponential improvements reproduce, to a relative error below $10^{-10}$, values saved from an earlier and separate run of the same computation. Our quadrature for the truncated-linear estimator matches an independent implementation to about $10^{-14}$. It also matches the limit derived in Appendix \ref{app: crossing}. A brute-force simulation from the raw formulas, with $10^7$ draws in the two specifications with non-negligible improvements, matches the computed risk of each of the four estimators to within two simulation standard errors. Unlike the loss of TSLS with three instruments, their losses have finite variance.

\bibliographystyle{plainnat}
\bibliography{references}

\end{document}